\documentclass[pdflatex,sn-mathphys]{sn-jnl}

\usepackage{amssymb}

\usepackage{amsmath}
\usepackage{bm}
\usepackage{graphicx}
\usepackage{dcolumn}
\usepackage{physics}
\usepackage{epstopdf}
\usepackage[caption=false]{subfig}
\usepackage{color}
\usepackage{float}

\usepackage{lineno}

\newcounter{bla}

\newcommand{\comment}[1]{}

\usepackage{listings}

\definecolor{codegreen}{rgb}{0,0.6,0}
\definecolor{codegray}{rgb}{0.5,0.5,0.5}
\definecolor{codepurple}{rgb}{0.58,0,0.82}
\definecolor{backcolour}{rgb}{0.95,0.95,0.92}
 
\lstdefinestyle{mystyle}{
    backgroundcolor=\color{backcolour},   
    commentstyle=\color{codegreen},
    keywordstyle=\color{magenta},
    numberstyle=\tiny\color{codegray},
    stringstyle=\color{codepurple},
    basicstyle=\footnotesize,
    breakatwhitespace=false,         
    breaklines=true,                 
    captionpos=b,                    
    keepspaces=true,                 
    numbers=left,                    
    numbersep=5pt,                  
    showspaces=false,                
    showstringspaces=false,
    showtabs=false,                  
    tabsize=2
}
\jyear{2021}%

\theoremstyle{thmstyleone}%
\theoremstyle{thmstyletwo}%

\theoremstyle{thmstylethree}%

\begin{document}

\title[Quantum multi-programming for Grover's search]{Quantum multi-programming for Grover's search}

\author[1]{\fnm{Gilchan} \sur{Park}}\email{gpark@bnl.gov}

\author[2]{\fnm{Kun} \sur{Zhang}}\email{kun.h.zhang@stonybrook.edu}

\author*[1]{\fnm{Kwangmin} \sur{Yu}}\email{kyu@bnl.gov}

\author[3]{\fnm{Vladimir} \sur{Korepin}}\email{vladimir.korepin@stonybrook.edu}

\affil[1]{\orgdiv{Computational Science Initiative}, \orgname{Brookhaven National Laboratory}, \orgaddress{\city{Upton}, \state{New York}, \postcode{11973},  \country{USA}}}

\affil[2]{\orgdiv{Department of Chemistry}, \orgname{Stony Brook University}, \orgaddress{\city{Stony Brook}, \state{New York}, \postcode{11794-3400},  \country{USA}}}

\affil[3]{\orgdiv{C.N. Yang Institute for Theoretical Physics}, \orgname{Stony Brook University}, \orgaddress{\city{Stony Brook}, \state{New York}, \postcode{11794-3840}, \country{USA}}}

\abstract{

Quantum multi-programming is a method utilizing contemporary noisy intermediate-scale quantum computers by executing multiple quantum circuits concurrently.
Despite early research on it, the research remains on quantum gates or small-size quantum algorithms without correlation.
In this paper, we propose a quantum multi-programming (QMP) algorithm for Grover's search.
Our algorithm decomposes Grover's algorithm by the partial diffusion operator and executes the decomposed circuits in parallel by QMP. We proved that this new algorithm increases the rotation angle of the Grover operator which, as a result, increases the success probability.
The new algorithm is implemented on IBM quantum computers and compared with the canonical Grover's algorithm and other variations of Grover's algorithms.
The empirical tests validate that our new algorithm outperforms other variations of Grover's algorithms as well as the canonical Grover's algorithm.

}

\keywords{Quantum Computing; NISQ Algorithm; Multi-Programming; Grover's Search.}

\maketitle


\section{Introduction}
\label{sec_intro}

In recent years, unprecedented development has taken place in a range of quantum computing.
In particular, significant progress has been made in building quantum computers by companies, such as IBM, Google, Intel, and Rigetti in superconducting architecture, and IonQ and Honeywell (Quantinuum) in trapped ion architecture, respectively.
The companies adapting superconducting architecture have announced around $50$ qubits quantum computers. In particular, IBM is operating a $127$-qubit quantum computer \cite{IBM_127q}, which is the largest number of  qubits now.
On the other hand, trapped ion quantum computers show better fidelity and full connectivity between qubits even though the number of qubits doesn't exceed $20$ now.
Regardless of the architecture, these Noisy-Intermediate Scale Quantum (NISQ) \cite{Preskill2018quantumcomputingin} computers suffer from several noises and errors such as measurements error, multi-qubit gate error, and even worse, short decoherence times.
Therefore, these contemporary quantum devices fail to produce reliable outputs for deep circuits with many gates today. 
Even though the quantum computer vendors are releasing better quantum computers in terms of Quantum Volume \cite{moll2018quantum, cross2019validating}, which is a method to measure and quantify the computational power of a quantum device, the quantum device still remains far not only from the quantum supremacy in practical problems, but also from implementing traditional quantum algorithms such as number factoring \cite{shor1994algorithms}, simulating quantum systems \cite{lloyd1996universal}, unstructured search \cite{groverQuantumMechanicsHelps1997}, and solving linear systems of equations \cite{harrow2009quantum}.
Furthermore, the availability of fault-tolerant quantum computers remains out of reach for now or decades away. 
Therefore, an imperative problem we face is efficiently using contemporary NISQ devices to achieve a quantum advantage.


In this vein, Variational Quantum Algorithms (VQAs), such as Quantum Approximate Optimization Algorithm (QAOA) \cite{farhi2014quantum}, Variation Quantum Eigensolver (VQE) \cite{peruzzo2014variational}, and Quantum Neural Networks \cite{farhi2018classification, altaisky2001quantum, beer2020training, cong2019quantum}, have drawn attention as a promising candidate to achieve quantum advantage on NISQ devices.
A different approach on NISQ devices is improving the traditional quantum algorithms such as Quantum Amplitude Estimation (QAE) \cite{aaronson2020quantum, suzuki2020amplitude, yu2020practical,  grinko2021iterative, rao2020quantum} and Grover's search \cite{zhang2020depth, zhang2021implementation,  zhang2022quantum} to work on NISQ devices.
These approaches are developing quantum algorithms suitable for NISQ devices. In particular, these algorithms struggle with long circuit depth. The former algorithms use classical computers in the middle, this undermines the quantum speed-up in spite of better accuracy and fidelity due to the shorter circuit depth.


Another approach to utilize contemporary noisy quantum computers more efficiently is by overlapping multiple circuits \cite{das2019case}.
This method, called Quantum Multi-Programming (QMP), utilizes the NISQ devices by executing multiple quantum circuits concurrently. 
The quantum circuits executed concurrently can be different and even have different circuit depths.
The main motivation of the QMP is that the number of qubits of NISQ computers is much greater  than the Quantum Volume (QV).
In particular, superconducting quantum computers have limited connectivity between qubits and low quantum volume compared to trapped ion quantum computers.
IBM's quantum volume, $V_{Q}$, definition is shown as follows:
\begin{equation*}
{\displaystyle \log _{2}V_{Q}={\underset {m}{\operatorname {arg\,max} }}\left\{\min \left[m,d(m)\right]\right\}}
\end{equation*}
where $m$ is the number of qubits and $d(m)$ is the depth of the model circuit with $m$-qubits \cite{cross2019validating}.
One depth of the model circuit is composed of a random permutation of the qubits involved in the test, followed by random two-qubit gates.
Therefore, the depth of the model circuit is greater than the circuit depth.
This expresses the maximum volume, width ($m$), and depth ($d(m)$), of the model circuits that can be implemented successfully on average by the computer. 
For example, $\it IBM\_Washington$ quantum computer has $127$ qubits with QV $64$.
This means the model circuit with $6$-qubits runs reliably at six depths of the model circuit on average on $\it{IBM\_Washington}$.
Since the quantum volume is the upper bound having a reliable output, this also means the $7$-qubit correlated system with the circuit depth of the model circuit fails to make reliable output even though $\it{IBM\_Washington}$ has $127$ qubits.
Another examples are that $\it{IBMQ\_Kolkata}$, $\it{IBMQ\_Montreal}$, and $\it{IBMQ\_Mumbai}$ have $27$ qubits and QV $128$. 
Other superconducting quantum computers share similar properties such as limited connectivity between qubits, and more qubits than the QV.
Therefore, it is imperative to use more efficiently the capacity of the contemporary superconducting quantum devices and the QMP is a recently suggested method for this purpose.


However, QMP on NISQ devices is not straightforward because QMP accompanies unfavorable impacts on the whole system such as measurement timing of the concurrent circuits \cite{das2019case} and crosstalk between different circuits \cite{ohkura2022simultaneous}.
In addition to those issues, efficient qubit mapping between the logical qubits and the physical qubits, and task scheduling are studied \cite{liu2021qucloud, niu2021enabling, niu2022parallel}, as well as the comparison of QMP between trapped ion and superconducting quantum devices \cite{niu2022multi}.
In spite of those preliminary studies for QMP, applying QMP to quantum algorithms is not developed well. 
We were able to find only one piece of literature that accelerated by QMP in Ref. \cite{resch2021accelerating}.

In this study, we apply QMP to Grover's search algorithm to enhance the performance of the algorithm on NISQ computers.
This algorithm decomposes Grover's algorithm by the partial diffusion operator and places the decomposed circuits in parallel.
This algorithm increases the rotation angle of the Grover operator so that, as a result, increases the success probability.
The detail of the algorithm is described in Sec. \ref{subsec:new_algorithm}.
This QMP variation of Grover's search outperforms Grover's search and several variations of it. The empirical results are described in Sec. \ref{sec:exp}.

\section{Grover's search algorithm}
\label{sec:GS}

Since the main contribution of this paper is a new variation of Grover's search algorithm and the key component of the new variation is the partial diffusion operator, we briefly review Grover's search algorithm and its variation, partial Grover's search algorithm.

\subsection{Original Grover's search algorithm}

The original Grover's search algorithm finds a target item (if there are multiple target items) in an unstructured database \cite{groverQuantumMechanicsHelps1997}. Suppose that the total number of items in the database is $N=2^n$. For each item, we can use a basis vector $\ket{i}$ in the Hilbert space $\mathcal H_2^{\otimes n}$ to represent each item. The power of quantum computation lies in quantum superposition. We can set the initial state as a uniform superposition of all the basis vectors, given by
\begin{equation}
\ket{s_n} = H^{\otimes n} \ket{0}^{\otimes n},
\end{equation}
with the single-qubit Hadamard gate $H$ \cite{NC10}. Note that all the Hadamard gates can be applied in parallel.

The target item, or the target state, can be recognized by the oracle. In Grover's search algorithm, the oracle is designed to reflect the sign of all the target states. It can be expressed as
\begin{equation}
O_T = 1\!\!1 - 2\sum_{t\in T} \ket{t} \bra{t},
\end{equation}
with the set of all target states $T$. Assume that the number of target states is $M=|T|$. Given the initial state $\ket{s_n}$, applying the oracle can not increase the probability of finding the target state. We would need an oracle-independent operation, called the diffusion operator. It is defined as
\begin{equation}
    D_n = 1\!\!1_{2^n} - 2 \ket{s_n} \bra{s_n}.
\end{equation}
The diffusion operator $D_n$ reflects the amplitudes of all states in terms of their average. Therefore the amplitudes of target and non-target states are mixed (diffused). 

Combining the oracle and the diffusion operator gives the Grover operator, denoted as
\begin{equation}
G_n = D_n O_T.
\end{equation}
Repeatedly applying the Grover operator on the initial state $\ket{s_n}$ would increase the amplitudes of target states. The probability of finding one of the target states after $j$ iteration of the Grover operator is
\begin{equation}
\label{eq:succ_prob_GS}
\abs{ \bra{t} G^j_n \ket{s_n} }^2 =
\sin^2\left((2j+1)\sin^{-1}\left(\sqrt{\frac{M}{N}}\right)\right).
\end{equation}
When $M/N\ll 1$, one can easily see that if $j$ approaches $\pi\sqrt{N/M}/4$, the success probability approaches 1. Then we only need $\mathcal O(\sqrt{N/M})$ numbers of oracle to find the target state. Compared to the classical exhaustive search with complexity $\mathcal O(N-M)$, Grover's search algorithm has a quadratic speedup.

\subsection{Partial Grover's search algorithm}
\label{subsec:divde_conquery}

Although the diffusion operator is oracle independent, it consumes computational resources. Specifically, the $n$-qubit diffusion operator is single-qubit-equivalent to the $n$-qubit Toffoli gate \cite{NC10}. In practice, the $n$-qubit Toffoli gate is hard to be implemented on NISQ computers when $n$ is large. The $n$-qubit Toffoli gate has to be decomposed into elementary single- and two-qubit gates. If the physical qubits are not connected, additional SWAP gates are also needed. 

To reduce the physical resources of the diffusion operators, Grover first proposed the partial diffusion operators in the quantum search algorithm \cite{groverTradeoffsQuantumSearch2002}. The local diffusion operator reflects the amplitudes in the subspace of the database. It has the form
\begin{equation}
D_m = \left(1\!\!1_{2^m}-2 \ket{s_m} \bra{s_m} \right)\otimes 1\!\!1_{2^{b}},
\end{equation}
with $b+m=n$. In other words, $b$ qubits are untouched by $D_m$. We also call it the local diffusion operator. The local diffusion operator $D_m$ is always easier to be implemented than the global diffusion operator $D_n$ on real computers with $m<n$. Correspondingly, we have the local Grove operator, given by
\begin{equation}
G_m = D_m O_T.
\end{equation}
Note that the local Grover operator has the same oracle as the global diffusion operator since the oracle is set once the problem to be solved is set. 

The local diffusion operator is also introduced to the partial search problem \cite{GR05,KG06}. It finds partial bits of the target state. It trades accuracy for speed. The quantum partial search algorithm is realized by the operator $G_nG^{j_2}_mG^{j_1}_n$. The number of iterations $j_1$ and $j_2$ is correlated in order to maximize the success probability. Meanwhile, one can also minimize the total number of oracles $j_1+j_2$ to give the optimal quantum partial search algorithm \cite{Korepin05}. Note that the quantum partial search algorithm for unevenly distributed multi-target states is studied in detail in Refs. \cite{choi2007quantum, zhang2018quantum}.

There is another way to design the search algorithm via the local diffusion operator. For simplicity, we assume that the database has a unique target state in the following, namely $M=1$. First, we can prepare the initial state $\ket{s_m} \otimes \ket{l}$ with the random bits $l\in \{0,1\}^{b}$. Correspondingly, the target state $\ket{t}$ is partitioned into $\ket{t}  = \ket{t_2} \otimes \ket{t_1}$. The partial target bit $t_1$ is also called the target block in the quantum partial search algorithm. The random guess has the success probability $1/2^b$ that $l=t_1$. If $l=t_1$, applying $G^j_m$ on $\ket{s_m} \otimes \ket{l}$ can find the rest of target bits $t_2$. The total success probability of finding the target state is
\begin{equation}
\label{eq:succ_prob_partial}
\frac{\sin^2 \left( (2j+1) \sin^{-1} \left( \sqrt{\frac{1}{2^m}} \right)  \right)}{2^b}.
\end{equation}
Since part of the target bits are found by quantum search, part of by random guess, such implementation is also called the hybrid classical-quantum search algorithm \cite{zhang2021implementation}. Not only do we need less number of oracles to achieve the maximal success probability, but all the global diffusion operators are replaced by the local diffusion operators. However, the drawback is obvious, namely the factor $1/2^b$ in the success probability. The multi-programming method aims to improve the success probability by removing $1/2^b$ from the success probability in Eq. (\ref{eq:succ_prob_partial}) by executing the partial search circuits in parallel. See \ref{subsec:new_algorithm} for details.

\section{Quantum Multi-Programming for Grover's search}
\label{sec:QMP-GS}

In spite of the unprecedented advance in quantum devices in recent years, NISQ computers are still susceptible to errors. 
Even though quantum error correction (QEC) code can be added to reduce the errors, the QEC code is another burden to NISQ computers.
To utilize contemporary noisy quantum computers more efficiently, the quantum multi-programming (QMP) is suggested in Ref. \cite{das2019case} and related studies followed \cite{ohkura2022simultaneous, liu2021qucloud, niu2021enabling, niu2022parallel, niu2022multi}.
The main motivation of the QMP is that the number of qubits of NISQ computers is much greater  than the quantum volume (QV).
QMP fills the gap between relatively many qubits and relatively low QV of NISQ devices by executing multiple quantum circuits concurrently, enhancing the throughput and utilization of NISQ devices.
After reviewing several issues in QMP briefly, we describe our new algorithm in Sec. \ref{subsec:new_algorithm}.

\subsection{Crosstalk between circuits}
\label{sec:Crosstalk}

The effect of crosstalk between qubits is studied in detail by P. Murali et al in \cite{murali2020software}. They suggested several rules to mitigate crosstalk between qubits.
The relation between the number of physical buffers and the error rate was studied by Ohkura et al \cite{ohkura2022simultaneous}.
They introduced a physical buffer, which is the number of idle qubits between quantum circuits.
They tested a different number of controlled-X (CX) gates with a different number of physical buffers.
In the test, they used the metric, probability of successful trial (PST) introduced in Ref. \cite{tannu2019not} as follows.

\begin{equation*}
PST = \frac{\textup{Number of successful trials}}{\textup{Total number of trials}}
\end{equation*}
\linebreak[1.5]

Ohkura et al.~\cite{ohkura2022simultaneous} concluded that only $1$ physical buffer is sufficient until $30$ CX gates because there is no PST decrease compared to a single circuit of CX gates when physical buffers $1$, $2$, and $3$ are inserted between the CX gates.
Also, their experiments show that the PST does not decrease even though there is no physical buffer between the quantum circuits when only $10$ CX gates are used.
Therefore, we use one physical buffer for our implementations.

\subsection{Other issues in QMP}

One crucial issue of QMP is measurement timing.
When quantum circuits having different circuit depths run concurrently, the measurement timing influences the shortest circuit. This effect was studied in Ref. \cite{das2019case, ohkura2021crosstalk}. In these studies, they suggested delaying shorter circuits to align the measurement timing. In our case, since the circuits in QMP have the same depth, we do not need to consider the measurement timing.

Another important issue in QMP is efficient physical qubit mapping. Systematic qubit mapping algorithms are developed \cite{das2019case, liu2021qucloud, niu2022parallel, niu2021enabling}.
Even though these qubit mapping algorithms are also important for ordinary quantum circuits (Not QMP), efficient qubit mapping is more crucial in QMP because QMP involves more qubits.

\subsection{New algorithm}
\label{subsec:new_algorithm}

\begin{figure}[t!]
\centering
\includegraphics[width=1\textwidth]{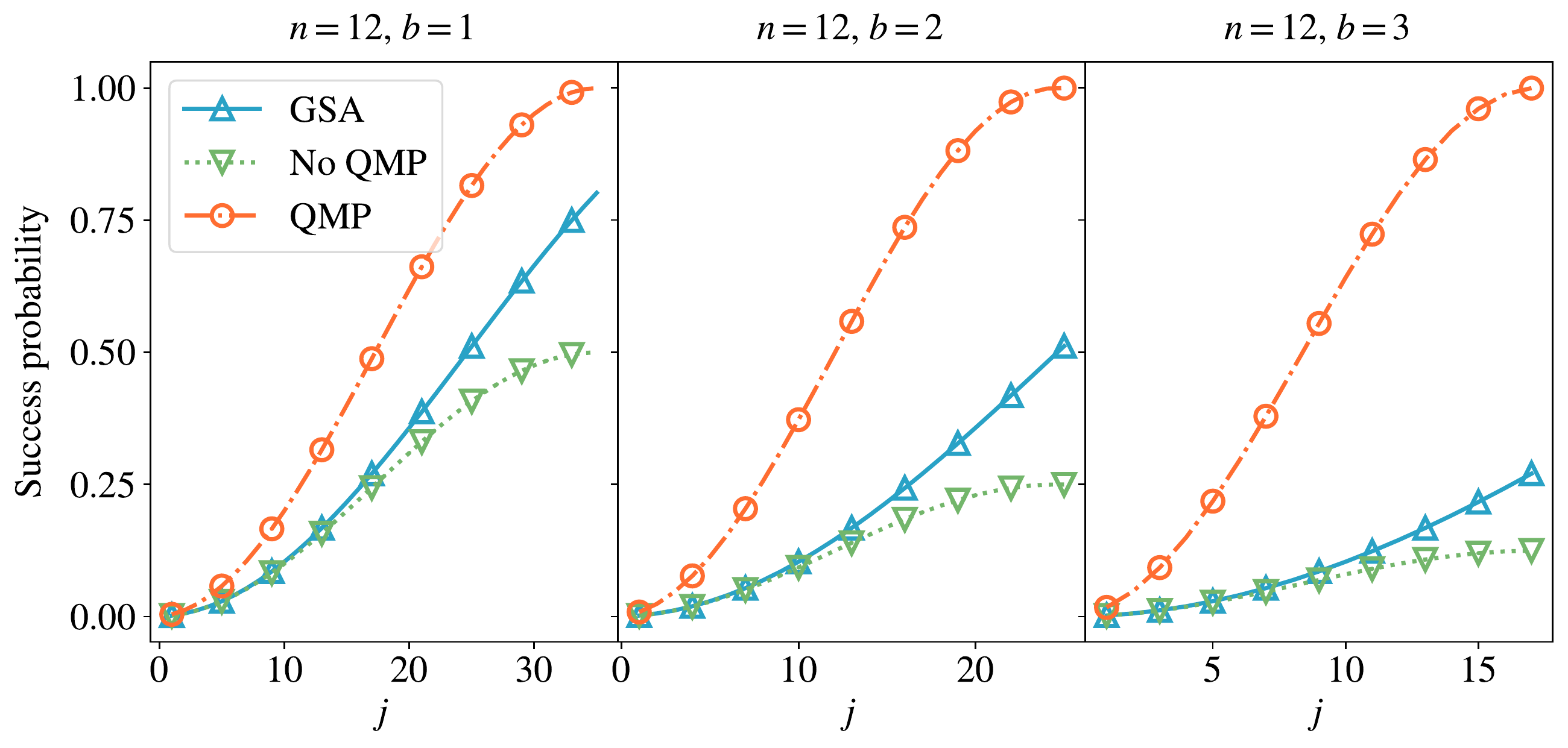}
\caption{Plots of success probabilities of three methods. This figure shows the success probabilities of Grover's search algorithm (GSA), partial Grover's search algorithm without multi-programming (No QMP) and quantum multi-programming search algorithm (QMP) with respect to $j$-times repeated Grover's operators in the GSA case or partial diffusion operators in the No QMP and QMP cases. These plots use one target out of $12$-qubit search domain ($4096$ items) with $b=1, 2$, and $3$. The value of $b$ represents the number of qubits for preset value in the partial Grover's search algorithm (No QMP). Therefore, we have $2^b$ blocks in the No QMP case, and $2^b$ circuits in the QMP search algorithm.}
\label{fig:ex_probability}
\end{figure}

The key idea of our new algorithm is placing the partial search algorithm in Sec. \ref{subsec:divde_conquery} in parallel through QMP.
Instead of a random choice of one block in the algorithm in Sec. \ref{subsec:divde_conquery}, we place all the cases on one quantum computer.
When we have $n$-qubit search domain and apply $m$-qubit ($m<n$) partial search, we have $b( = n-m)$-qubit guesses in the partial search algorithm.
The success probability of finding the target states (solutions) is

\begin{equation*}
   \frac{\sin^2 \left( (2j+1) \sin^{-1} \left( \sqrt{\frac{1}{2^m}} \right)  \right)}{2^{b}}
\end{equation*}

\noindent
as described in Eq. (\ref{eq:succ_prob_partial}).
When we place $B (=2^b)$-circuits in the QMP, we have $B$ times better success probability as follows:
\begin{equation}
\label{eq:QMPGS_succ_prob}
   \sin^2 \left( (2j+1) \sin^{-1} \left( \sqrt{\frac{1}{2^m}} \right)  \right).
\end{equation}
The canonical Grover's algorithm has the success probability as follows:
\begin{equation}
\label{eq:GS_succ_prob}
   \sin^2 \left( (2j+1) \sin^{-1} \left( \sqrt{\frac{1}{N}} \right)  \right).
\end{equation}

\noindent
where we have $N = 2^n$.
When we rewrite Eq. (\ref{eq:QMPGS_succ_prob}) in terms of $B$ and $N$, we have
\begin{equation}
\label{eq:QMPGS_succ_prob_reform}
   \sin^2 \left( (2j+1) \sin^{-1} \left( \sqrt{B} \sqrt{\frac{1}{N}} \right)  \right).
\end{equation}
Since $\sin^{-1} \left( \sqrt{\frac{1}{N}} \right)$ in Eq. (\ref{eq:GS_succ_prob}) is the rotation angle of the Grover operator, we have the rotation angle of our QMP Grover's search algorithm as follows:
\begin{equation}
\label{eq:QMPGS_angle}
    \theta_{MP} = \sin^{-1} \left( \sqrt{B} \sqrt{\frac{1}{N}} \right).
\end{equation}
On the other hand, the quantum states of the blocks which do not include the targets remain amplitude $\sqrt{B/N}$.

When we have $0 < \sqrt{\frac{1}{N}} < \sqrt{B} \sqrt{\frac{1}{N}} < 1$, the arcsine function is an increasing function. Therefore, we have $\theta_{MP}$ which is greater than the rotation angle of the canonical Grover's algorithm.
The greater rotation angle increases the efficiency of the search algorithm.

Figure \ref{fig:ex_probability} shows the success probabilities of Grover's search algorithm (GSA), partial Grover's search algorithm without multi-programming (No QMP), and quantum multi-programming search algorithm (QMP) with respect to $j$-times repeated Grover's operators in the GSA case and partial diffusion operators in the No QMP and QMP cases, respectively. These plots use one target out of $12$-qubit search domain ($4096$ items) with $b=1, 2$, and $3$. The value of $b$ represents the number of qubits for preset value in the partial Grover's search algorithm (No QMP). Therefore, we have $2^b$ blocks in the No QMP case, and $2^b$ circuits in the QMP search algorithm.
These plots compare the success probabilities of three quantum search algorithms in theory.
The success probabilities of GSA, No QMP, and QMP are shown in Eq. (\ref{eq:succ_prob_GS}),(\ref{eq:succ_prob_partial}), and (\ref{eq:QMPGS_succ_prob_reform}), respectively.
In this specific case, the GAS has the rotation angle of $\sin^{-1} \left( \frac{1}{\sqrt{2^{12}}} \right)$ ($\approx1.56256 \times 10^{-2}$ radian) regardless of $b$.
On the other hand, the QMP and the No QMP have the rotation angle of $\sin^{-1} \left( \sqrt{2^b} \frac{1}{\sqrt{2^{12}}} \right)$.
Therefore, the rotation angles are $\sin^{-1} \left( \frac{1}{\sqrt{2^{11}}} \right)$, $\sin^{-1} \left( \frac{1}{\sqrt{2^{10}}} \right)$, and $\sin^{-1} \left( \frac{1}{\sqrt{2^{9}}} \right)$ when we have $b=1, 2$, and $3$, respectively.
Hence, the success probabilities of the QMP and the No QMP increase as the $b$ increases in a fixed $j$.
Even though the success probability of the No QMP is lower than the success probability of the QMP because it is divided by $2^b$, it is still greater than the success probability of the GSA.
The efficiency of the No QMP over the GSA is detailed in Ref. \cite{zhang2020depth, zhang2021implementation}.
Also, these plots show that the QMP is much more efficient than the No QMP as we have greater $b$.
When we have $b=3$, the QMP needs $14$-times iteration of the partial Grover's operators to achieve $90 \%$ success probability while the GAS needs $40$-times iteration of the partial Grover's operators to achieve $90 \%$ success probability.

\subsection{Partial measurement}

The result of each quantum circuit in QMP is a partial measurement related to the circuit.
The partial measurement for each circuit is post-processed after measuring all circuits since the partial measurement probability is the sum of all other uninvolved qubits measurement probability.
A sample code extracting partial measurement sampling out of the whole measurement sampling is listed in Appendix \ref{app:meas}.
A figure-of-merit regarding measurement in QMP, Trial Reduction Factor (TRF), is introduced in Ref. \cite{das2019case}.
The criterion describes the ratio of the number of the circuit which executes individually (baseline) to the number of trials of the QMP circuit.
In our case, the sampling results of each circuit can be extracted from the whole measurement as described above, the TRF is $B$ (the number of circuits of the QMP).

\begin{figure}[t]
\centering
  \subfloat[Circuit diagram]{%
    \includegraphics[width=0.49\textwidth]{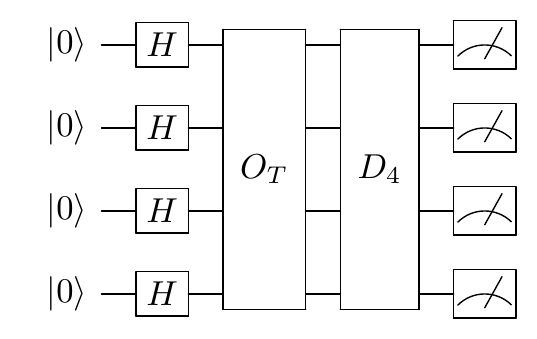}}
  ~ 
  \subfloat[Sampling histogram]{%
    \includegraphics[width=0.49\textwidth]{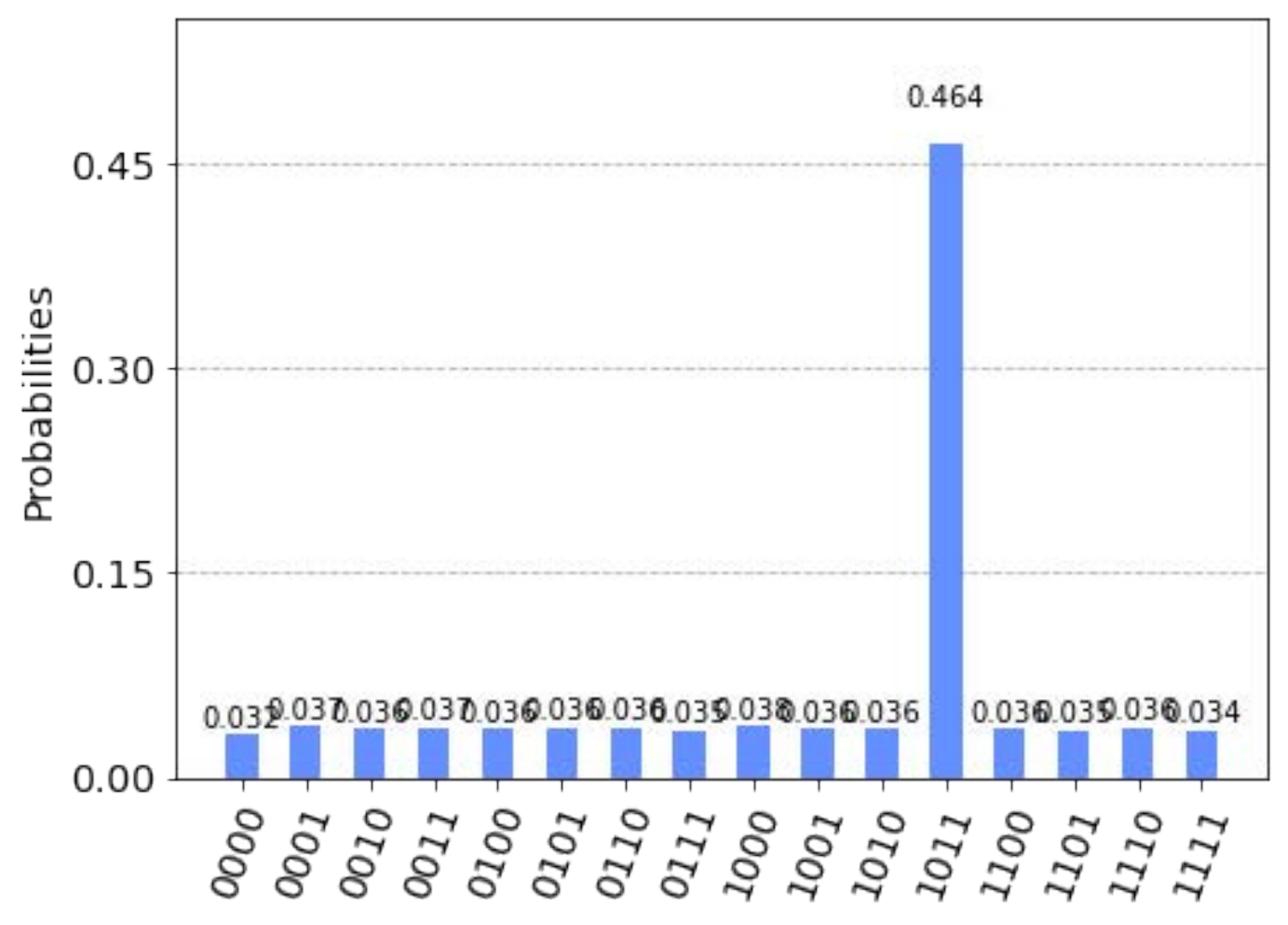}}
  \caption{The canonical Grover's search. This figure shows the quantum circuit diagram of the canonical Grover's search with target $\ket{1011}$ and its sampling histogram with $8192$ trials (shots).}
  \label{fig:qc_4q_no_qmp_sim}
\end{figure}

\subsection{Example}
\label{sec:example1}

\begin{figure}[t!]
\centering
\includegraphics[width=0.4\textwidth]{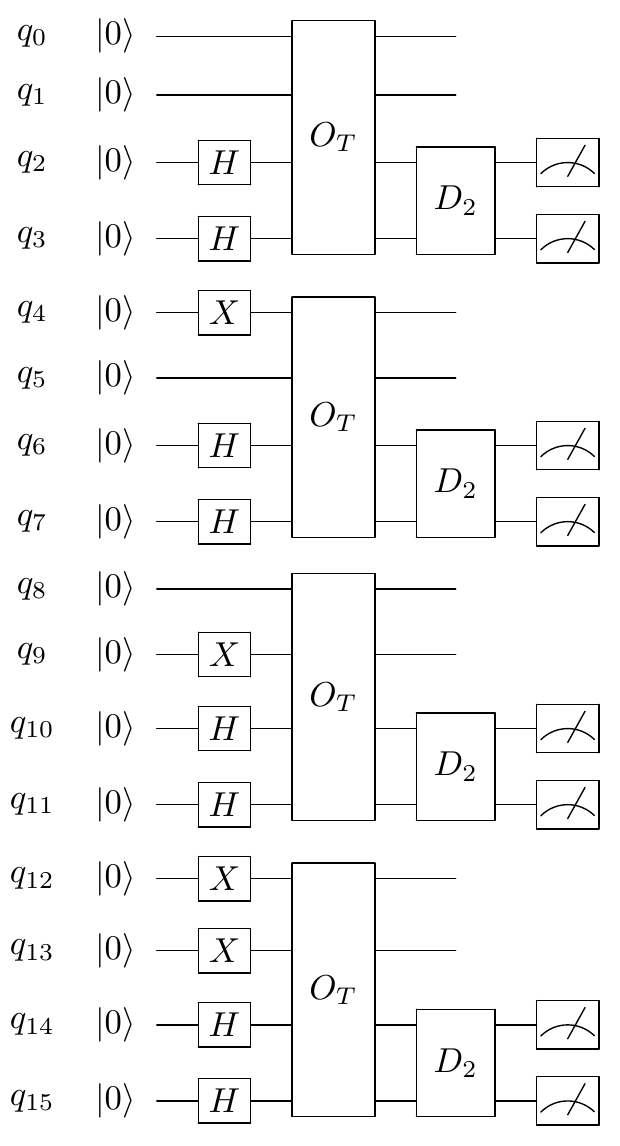}
\caption{Quantum circuit diagram of the QMP. This figure shows the quantum circuit diagram of the QMP for the circuit in Fig \ref{fig:qc_4q_no_qmp_sim}.}
\label{fig:qc_4q_qmp_sim}
\end{figure}

In this section, we explain our new algorithm with an example.
Even though this algorithm works with multiple search targets (solutions), this example has one target for simplicity. 
The multiple targets case is explained in Sec.~\ref{sec:multi_target} in detail.
In the bit ordering convention, the rightmost and the uppermost qubits are the least significant bit (LSB) in the bra-ket notation and the quantum circuit, respectively.
In this example, we have the target $\ket{1011}$ ($\ket{MSB \leftrightarrow LSB}$).
Figure \ref{fig:qc_4q_no_qmp_sim} shows a diagram for the quantum circuit in Qiskit of the canonical Grover's search.
The Grover's operator is applied once.
Therefore, the success probability of this search is $\sin^2 \left( 3 ~ \sin^{-1} \left(  \sqrt{\frac{1}{2^4}} \right)  \right) \approx 0.472656$.
The simulation result of $8192$ trials (shots) is shown in Fig. \ref{fig:qc_4q_no_qmp_sim}.
The target sampling probability is $0.464$ which is consistent with the theoretical success probability.

\begin{figure}[t!]
\centering
  \subfloat[Whole measurement]{%
    \includegraphics[width=0.49\textwidth]{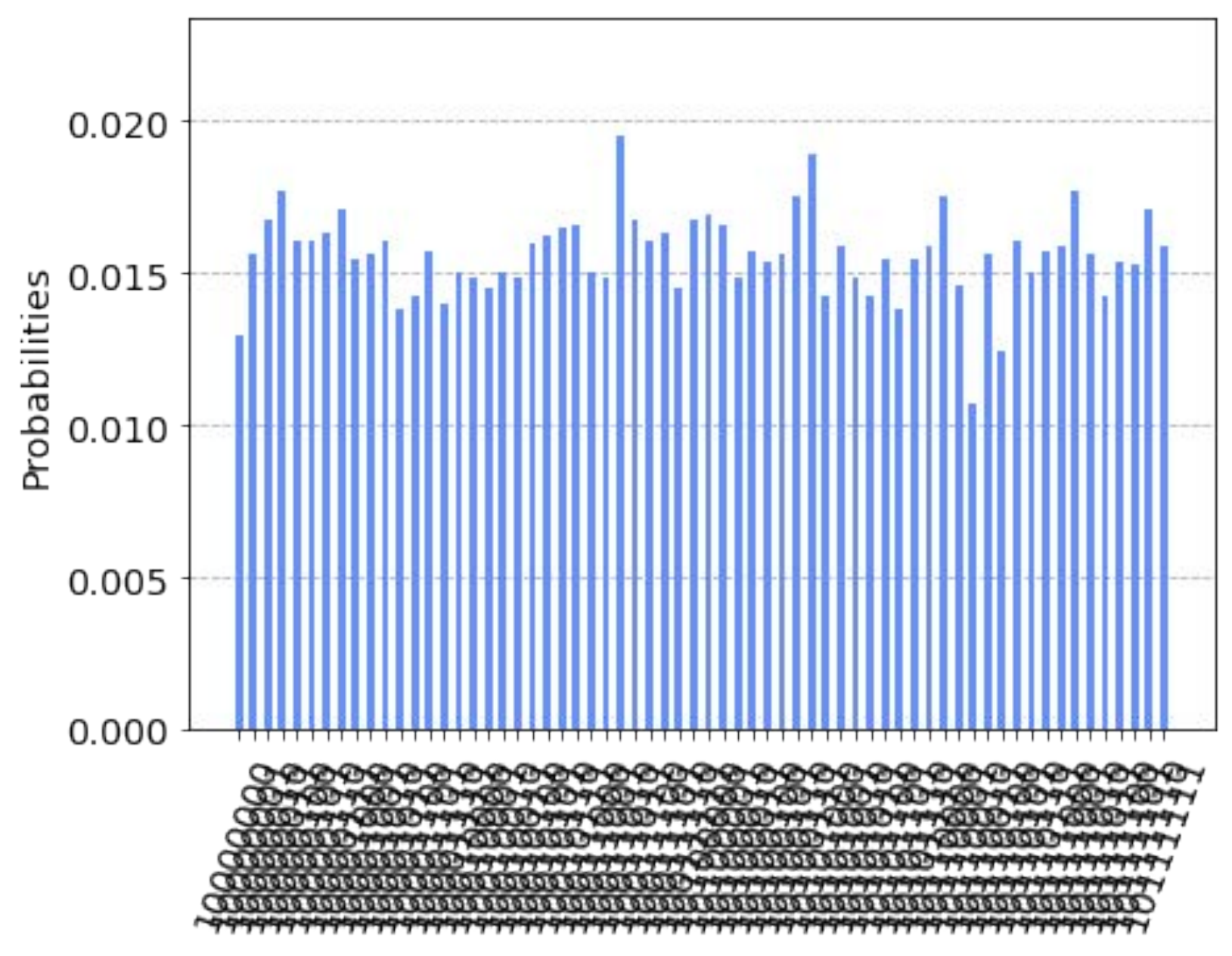}}
    ~~
  \subfloat[Partial measurement]{%
    \includegraphics[width=0.49\textwidth]{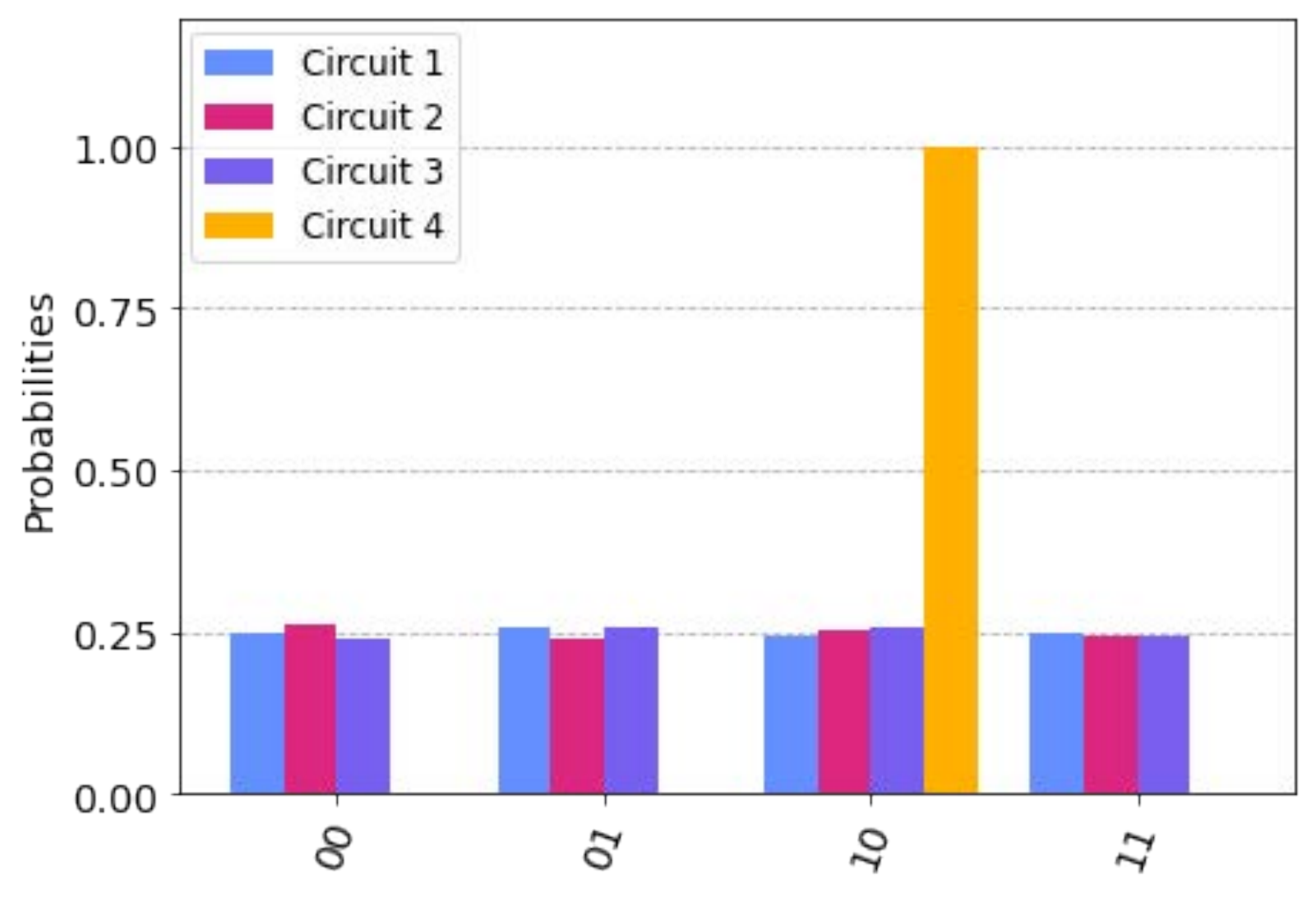}}
  \caption{The sampling histogram of the circuit of Fig. \ref{fig:qc_4q_qmp_sim} with $8192$ trials (shots). Figure (a) shows the result of the whole measurement of the circuit in Fig. \ref{fig:qc_4q_qmp_sim}. The result of Circuit 1 is the sampling of measurement of qubit $q_2$ and $q_3$. This partial measurement in Fig. (b) is estimated by the algorithm listed in Appendix \ref{app:meas} based on the sampling result shown in Fig. (a).}
  \label{fig:hist_4q_no_qmp_sim_partial}
\end{figure}

The QMP circuit diagram is shown in Fig. \ref{fig:qc_4q_qmp_sim}. The diffusion operators are partial diffusion operators. The QMP circuit is composed of four partial search circuits with initial guesses of the first and the second qubits.
For example, the first circuit is placed on qubits from $q_0$ to $q_3$.
The first and the second qubit, $q_0$ and $q_1$, are initialized by $\ket{00}$.
The second circuit is placed on qubits from $q_4$ to $q_7$.
The first and the second qubit of the second circuit, $q_4$ and $q_5$, of the circuit, are initialized by $\ket{01}$.
In each circuit, since the first and the second qubits are initialized with specific bits and the third and the fourth qubits are superpositioned with the Hadamard gates, only the third and the fourth qubits are measured.

From the sampling result of $8192$ trials (shots) shown in Fig. \ref{fig:hist_4q_no_qmp_sim_partial} (a), we extract the partial measurement sampling by using the code in Appendix \ref{app:meas}.
The sampling results of each circuit of the QMP are shown in Fig. \ref{fig:hist_4q_no_qmp_sim_partial} (b).
The result of Circuit 1 is the sampling of measurement of qubit $q_2$ and $q_3$.
The result of Circuit 2 is the sampling of measurement of qubit $q_5$ and $q_6$.
Since the first and the second qubits from the right (LSB) of the target $\ket{1011}$ are $\ket{11}$, the fourth circuit includes the target and all other circuits do not have the target.
Therefore, the partial diffusion operators of the first, the second, and the third circuits do not amplify the amplitude of the initial quantum states after the Hadamard gates and they have a uniform distribution of $25 \%$ probability as shown in Fig. \ref{fig:hist_4q_no_qmp_sim_partial}.
On the other hand, since the fourth circuit includes the target, the partial diffusion operator amplifies the quantum states of the target.
The success probability of the fourth circuit is $\sin^2 \left( 3 ~ \sin^{-1} \left(  \sqrt{2^2} \sqrt{\frac{1}{2^4}} \right)  \right) = 1.0$ which is consistent with the  histogram of the circuit $4$ in Fig. \ref{fig:hist_4q_no_qmp_sim_partial} and Eq. (\ref{eq:QMPGS_succ_prob_reform}).

\subsection{Extension to multiple targets}
\label{sec:multi_target}

\begin{figure}[t!]
\centering
  \subfloat[The canonical GS]{%
    \includegraphics[width=0.49\textwidth]{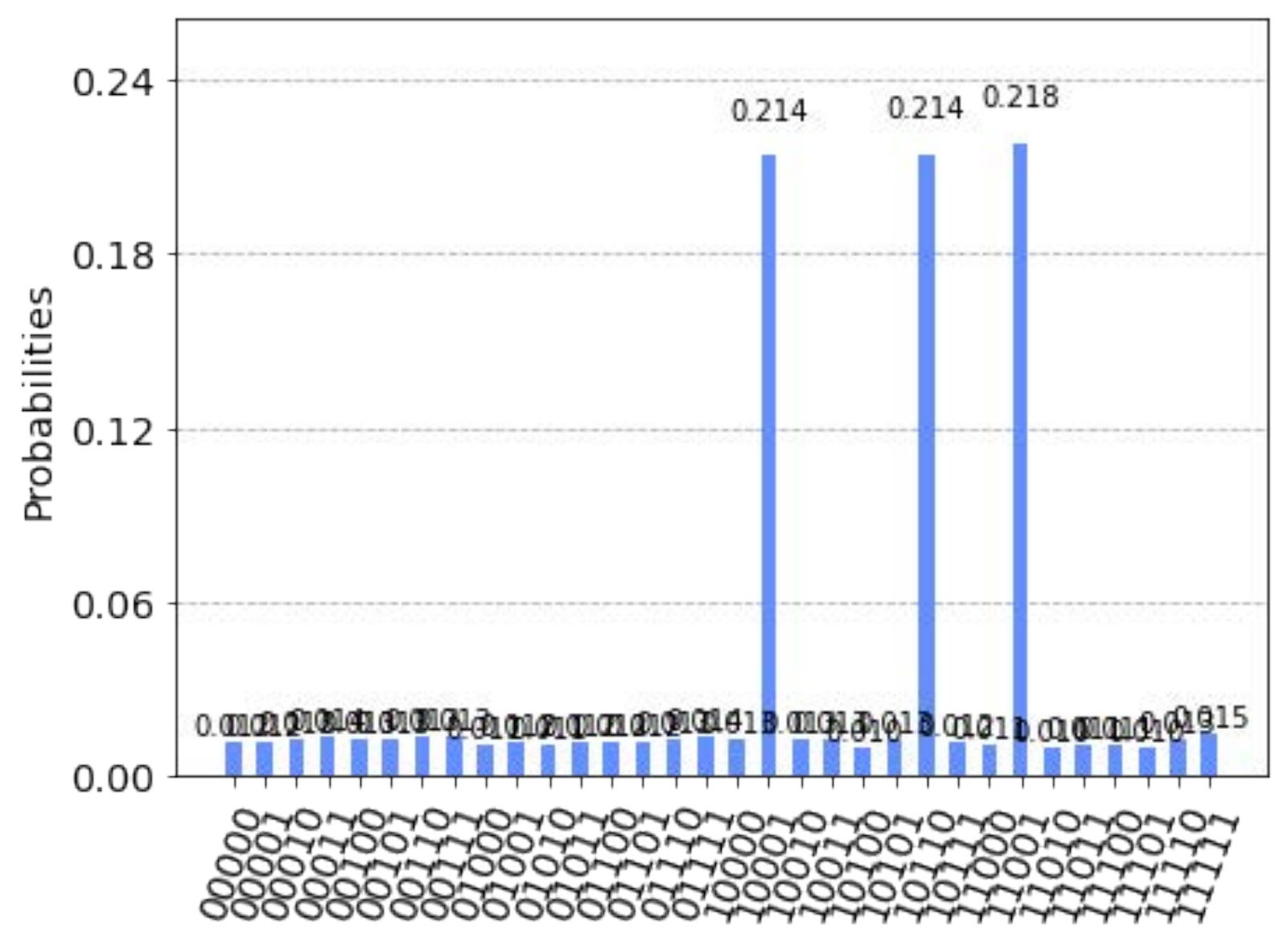}}
    ~~
  \subfloat[The QMP GS]{%
    \includegraphics[width=0.49\textwidth]{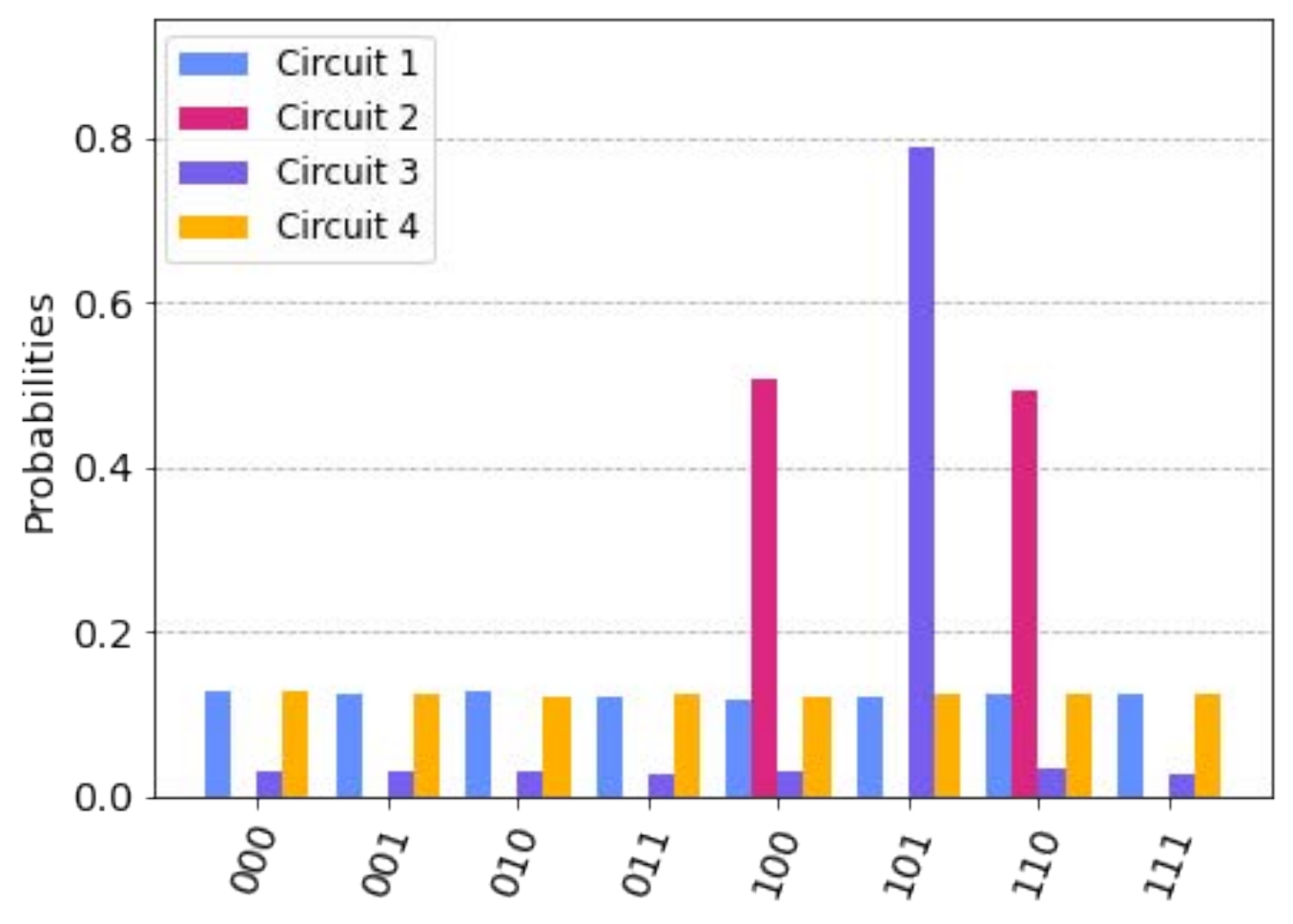}}
  \caption{The result of multiple targets. The simulation results of the example of three targets out of five qubit search domains with $8192$ trials (shots). Figure (a) shows the result of the canonical Grover's search and Fig. (b) shows the result of the QMP Grover's search.}
  \label{fig:hist_5q_no_qmp_sim}
\end{figure}

The quantum partial search algorithm for multi-target states is studied in detail in Refs. \cite{choi2007quantum, zhang2018quantum} including the uneven distributions of the targets over the partial blocks.
Based on the study of the multi-target case of the partial Grover's search, we extend the QMP Grover's search and the success probability of the QMP Grover's search in Eq. (\ref{eq:QMPGS_succ_prob_reform}) is extended to multiple target case as follows:
\begin{equation}
\label{eq:QMPGS_succ_prob_reform_multi}
   \sin^2 \left( (2j+1) \sin^{-1} \left( \sqrt{B} \sqrt{\frac{M_i}{N}} \right)  \right)
\end{equation}
where $M_i$ is the number of targets in $i$-th block and the total number of targets, $M= \displaystyle\sum_{i=1}^{B} M_i$.
This success probability per block depends on the number of targets belonging to the block.
As with the single target case, the quantum states of the blocks which do not include the targets remain amplitude $\sqrt{B/N}$.

We demonstrate the multiple target case of the QMP Grover's search with three targets, $\ket{10110}$, $\ket{10001}$, and $\ket{11001}$.
Similar to Sec. \ref{sec:example1}, the search domain is group by four blocks, $\ket{*** ~ 0 0}$, $\ket{*** ~ 0 1}$, $\ket{*** ~ 1 0}$, and $\ket{*** ~ 1  1}$ (number by block $1$, $2$, $3$, and $4$, respectively), with the same bit ordering convention.
Therefore, the target $\ket{10110}$ belongs to block $3$ the targets  $\ket{10001}$ and $\ket{11001}$ belong to block $2$, respectively.

Figure \ref{fig:hist_5q_no_qmp_sim} (a) is the sampling histogram of the canonical Grover's search and Fig. \ref{fig:hist_5q_no_qmp_sim} (b) is the result of our new algorithm, QMP Grover's search.
In those simulations, the Grover operators are applied once.
The success probability of the canonical GS is $\sin^2 \left( 3 ~ \sin^{-1} \left(  \sqrt{\frac{3}{2^5}} \right)  \right) \approx 0.645996$ (refer to Eq. (\ref{eq:succ_prob_GS})) and Fig. \ref{fig:hist_5q_no_qmp_sim} (a) shows that the sum of the three probabilities of the three targets is $0.642$.

In the QMP Grover's search as shown Fig. \ref{fig:hist_5q_no_qmp_sim} (b), the success probabilities depend on the number of targets in the block.
Since the block $3$ (or the circuit $3$) includes one target, $\ket{10110}$, its success probability is $\sin^2 \left( 3 ~ \sin^{-1} \left(\sqrt{4} \sqrt{\frac{1}{2^5}} \right)  \right) \approx 0.78125$ (refer to Eq. (\ref{eq:QMPGS_succ_prob_reform})).
The block $2$ (or the circuit $2$) includes two targets, $\ket{10001}$ and $\ket{11001}$. Therefore, its success probability is $\sin^2 \left( 3 ~ \sin^{-1} \left(\sqrt{4} \sqrt{\frac{2}{2^5}} \right)  \right) = 1.0$.
The histogram of the circuit $2$ shows that the two targets have about $0.5$ probabilities and the probabilities of non-target are zero.
Since the circuit $1$ and $4$ do not include the targets, each quantum state remains in its initial state $\sqrt{1/8}$ as shown in Fig. \ref{fig:hist_5q_no_qmp_sim} (b).

\section{Empirical tests}
\label{sec:exp}

This section describes and compares the result of the five-qubit quantum search problem with one target on several IBM quantum computers.
Our QMP algorithm for Grover's search is compared with the partial Grover's search algorithm of the $3$-qubit guess and the $2$-qubit guess cases.
Before reviewing the experiment results, we briefly describe the metrics for the comparison.

\subsection{Experimental metrics}
\label{subsec:test_metrics}

We use the following two criteria to compare the different quantum circuit implementations.
The success probability is a criterion for Grover's search algorithm.
We suggest the expected quantum circuit volume to compare the efficiency of the implementations.

\subsubsection{Success probability}

The success probability depends on the ratio of the number of targets (solutions) to the size of the search domain and the number of the Grover operator. The success probabilities for Grover's search, the partial Grover's search, and our new QMP search algorithm are described in Eq. (\ref{eq:succ_prob_GS}), (\ref{eq:succ_prob_partial}), and (\ref{eq:QMPGS_succ_prob_reform}), respectively.

\subsubsection{Expected quantum circuit volume}
We propose a metric, Quantum Circuit Volume (QCV), for the usage of quantum computer resources of a quantum circuit as follows:

\begin{equation}
\label{eq:QCV}
    QCV (O) = nq (O) \times d(O).
\end{equation}
\noindent
where $O$ is a quantum circuit implementation of a quantum algorithm, $nq(O)$ is the total number of qubits used to implement $O$, and $d(O)$ is the circuit depth of the implementation $O$.
The QCV measures how much quantum computer resource, qubits (space), and circuit depth (time), are used.

Also, we propose a comparison metric, Expected Quantum Circuit Volume (EQCV), as an extension of the expected depth in Ref. \cite{zhang2021implementation, zhang2022quantum}.
The expected depth is a normalization of the circuit depth by the success probability.
It is the ratio of the circuit depth to the success probability.
It describes the efficiency of the search algorithm or the implementation.

However, the expected circuit depth is not appropriate to compare the resource efficiency between the canonical Grover's search algorithm and its QMP variation because the QMP uses several times more qubits than the canonical Grover's search. 
Hence we define EQCV as follows:

\begin{equation}
\label{eq:EQCV}
    EQCV (O) = \frac{QCV (O)}{\textup{Success Probability of}~O}.
\end{equation}

The circuit depth in the expected depth is replaced by the QCV in the EQCV.
We use the EQCV instead of the expected depth to compare the efficiency of the algorithms in this section.
In this experiment, our QMP implementation has multiple times QCV than the partial Grover's search implementation.
For example, when two qubits are fixed for guess in the partial Grover's search, the QMP of it has four circuits. So the QCV of the QMP is four times greater than the QCV of the partial Grover's search.

\subsection{Quantum circuit implementation}
\label{sec:implementation}

This section describes the quantum circuit implementation which is used in these experiments.
We use the same bit order convention with the convention in Sec. \ref{sec:example1} and \ref{sec:multi_target}.
In this bit ordering convention, the rightmost and the uppermost qubits are the least significant bit (LSB) in the bra-ket notation and the quantum circuit, respectively.

\begin{itemize}
    \item D5M5: the original five-qubit Grover's search algorithm with one iteration of the Grover operator (with a five-qubit diffusion operator). Here D5 suggests the five-qubit diffusion operator; M5 suggests performing the five-qubit measurements. It has the quantum circuit diagram
    \begin{center}
    \includegraphics[height=12em]{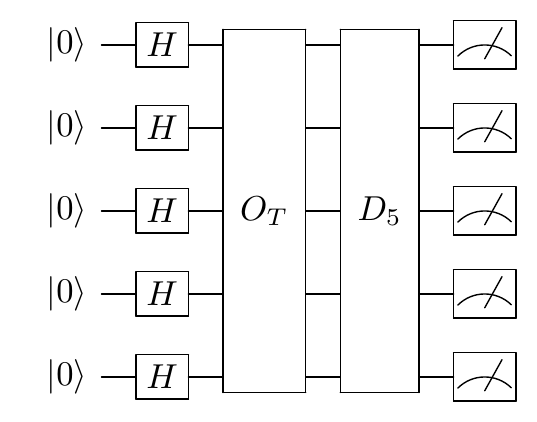}
    \end{center}
    
    \item D5D5M5: the original five-qubit Grover's search algorithm with two iterations of the Grover operator. It has the quantum circuit diagram
    \begin{center}
    \includegraphics[height=12em]{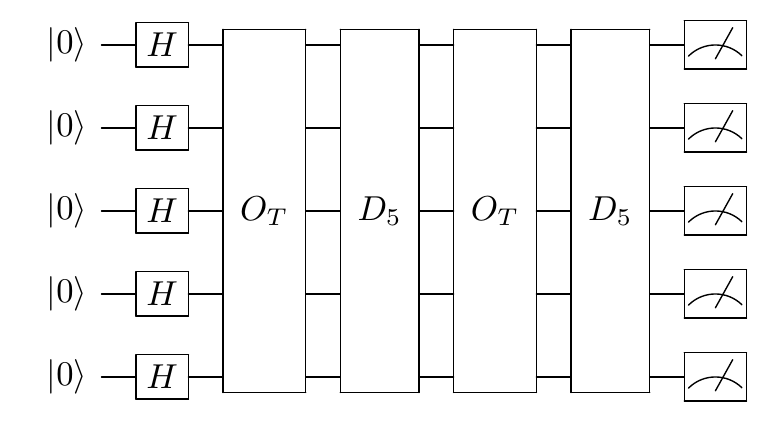}
    \end{center}
    
    \item G2D3M3: randomly guess the values of two qubits (denoted as G2) then apply the local Grover operator with a three-qubit diffusion operator and measure the three qubits. It has the quantum circuit diagram
    \begin{center}
    \includegraphics[height=12em]{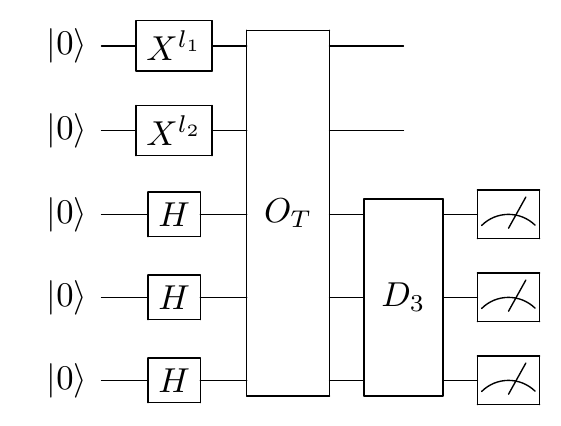}
    \end{center}
    Here $X$ is the Pauli-$x$ gate and $l_1,l_2\in\{0,1\}$ are randomly chosen. 
    
    \item G3D2M2: randomly guess the values of three qubits then apply the local Grover operator with a two-qubit diffusion operator and measure the two qubits. It has the quantum circuit diagram 
    \begin{center}
    \includegraphics[height=12em]{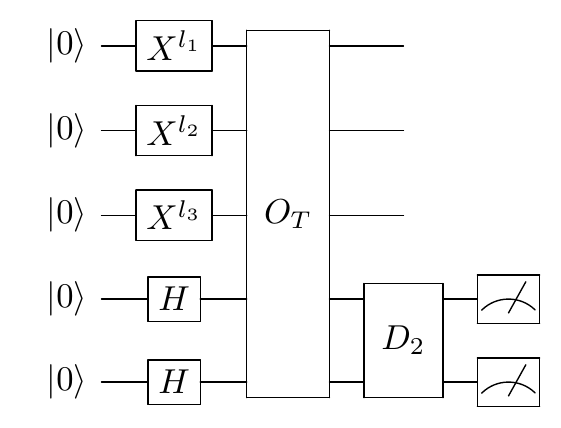}
    \end{center}
\end{itemize}

An important module of those above implementations is the five-qubit Toffoli gate (four-qubit-controlled X gate).
This five-qubit Toffoli is used for the five-qubit oracle ($O_T$) and five-qubit diffusion operator ($D_5$).
For the five-qubit Toffoli gate, we followed the implementation in Ref. \cite{zhang2022quantum}.
The implementation details and experimental results of partial Grover’s search are detailed in Refs. \cite{zhang2021implementation, zhang2022quantum}. In particular, the 5-qubit cases are detailed in Sec. 4.5 and Table 13 in Ref. \cite{zhang2021implementation}.

\subsection{Experimental setting}
\label{sec:setting}

We conducted the experiments of the designed quantum circuits on the two IBM quantum computers including $\it{IBM\_Washington}$ ($127$ qubits, $64$ QV) and $\it IBMQ\_Brooklyn$ ($65$ qubits and $32$ QV). We have the target $\ket{10110}$ (\textit{MSB} $\leftrightarrow$ \textit{LSB}) as the solution, and the G2D3M3 and G3D2M2 circuits take the initial values of two and three LSB qubits and measure the rest qubits, respectively. 
In order to efficiently implement the  five-qubit Toffoli gate, one ancillary qubit is added \cite{zhang2022quantum}. Thus the number of qubits in a circuit is six. 

To prepare the quantum circuit for the IBM quantum computers, first, we constructed the whole multi-programming circuit like the circuit diagram in Fig. \ref{fig:qc_4q_qmp_sim}. After that, we applied the Qiskit transpiler to the whole QMP circuit with the optimization level of $3$ for each target IBM quantum computer.
To run the experiments of G2D3M3 and G3D2M2 circuits, $24$ ($6$: qubits in a circuit $\times$ 4: circuits (number of cases of guess qubit values - 00,01,10,11)) and 48 qubits are used, respectively. 
The logical qubits are aligned for consecutive logical qubits to form a circuit in the whole QMP circuit from index $0$. For example, the first circuit has from qubit $0$ to qubit $5$ and the second circuit has from qubit $6$ to qubit $11$.
We tested two different QMP circuits for G2D3M3 and G3D2M2.
One is represented by ``QMP (default)" and the another is represented by ``QMP (manual)''. The former used the Qiskit transpiler without additional options except for the optimization level and the latter added the initial qubit mapping (``initial\_layout'' option) in addition to the optimization level.
The initial qubit layout for QMP (manual) is designed to guarantee at least one (physical) buffer qubit between each circuit in the QMP circuit as recommended in Sec. \ref{sec:Crosstalk}, and for logical qubits in a circuit to be allocated physically adjacent as well as better connectivity, and fewer errors of qubits and connections.
In QMP (manual), we arbitrarily selected the physical qubits to satisfy the above conditions of QMP (manual).
Figure. \ref{fig:G2D3M3_brooklyn_layout}, \ref{fig:G2D3M3_washington_layout}, and \ref{fig:G3D2M2_washington_layout} show the qubit mapping layouts on $\it IBM\_Washington$ and $\it IBMQ\_Brooklyn$.
Also, the figures show the difference between QMP (default) and QMP (manual).
In the G3D2M2 case, QMP (manual) needs $48$ qubits and additional buffer qubits. Due to the lack of qubit layout mapping space satisfying the conditions of QMP (manual)  in $\it IBMQ\_Brooklyn$ (Refer to Fig. \ref{fig:G2D3M3_brooklyn_layout} for $\it IBMQ\_Brooklyn$ layout), G3D2M2 was tested only on $\it IBM\_Washington$.

For the comparison, we also ran the partial Grover's search without QMP (``No QMP''), which operated on a single circuit with the solution values for guess qubits while the QMP circuit includes all circuits for possible bit strings as explained in Sec. \ref{sec:QMP-GS}.
More details with respect to the partial Grover's search without QMP is discussed in Sec. \ref{subsec:divde_conquery} and Ref. \cite{zhang2020depth, zhang2021implementation, zhang2022quantum}.
No QMP circuits used the same initial qubit mapping with QMP (default). That is, they are transpiled with optimization level $3$ and without the initial qubit mapping layout (with the ``initial\_layout=None'' option).
The original five-qubit Grover's search algorithm (GSA) also ran on a single circuit on $\it IBM\_Washington$. 
They were transpiled in the same manner as QMP (manual).

The experiments were performed with $8192$ trials (shots) and 30 running repetitions, and we used the average value of the $30$ repetitions for the height of the histograms with the standard deviation error bar in Sec. \ref{sec:results}.

\subsection{Experiment results}
\label{sec:results}

In this section, we discuss the experimental results.
The circuit depth and qubit mapping layout in this discussion are listed in Appendix \ref{app:no_qmp_orig_grover_layout}.

\subsubsection{Success probability}

\begin{figure}[t!]
\centering
\includegraphics[width=0.99\textwidth]{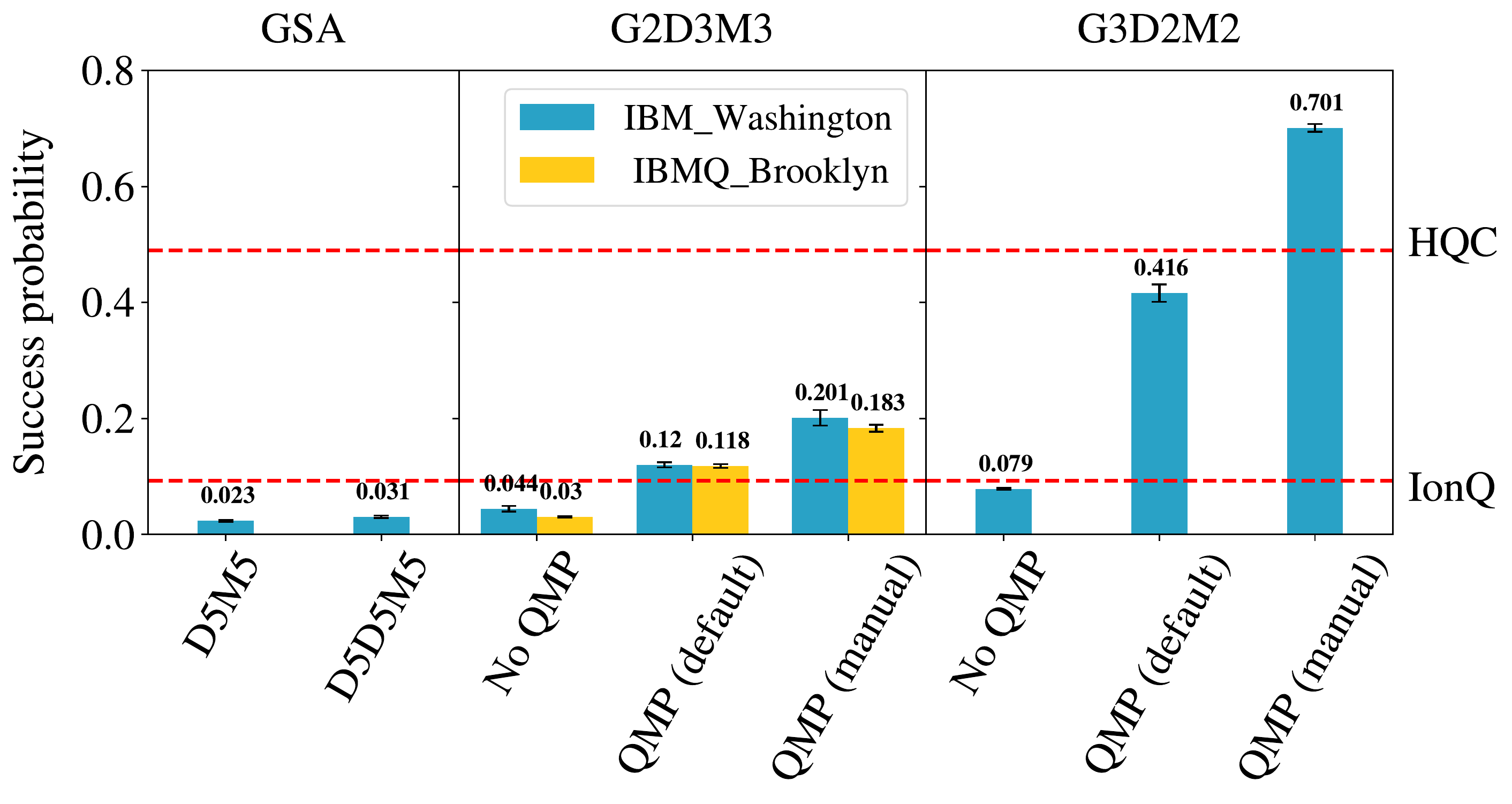}
\caption{Success probabilities of the experimental results. Refer to Sec. \ref{sec:implementation} and Sec. \ref{sec:setting} for each experimental case. The upper dotted red line and the lower dotted red line are the success probabilities of Honeywell quantum computer (HQC) ($49.0 \%$) and IonQ quantum computer ($9.3 \%$) from Ref. \cite{zhang2022quantum}. The experiments were performed with $8192$ trials (shots) and 30 running repetitions, and we used the average value of the $30$ repetitions for the height of the histograms with the standard deviation error bar. } 
\label{fig:test_probability}
\end{figure}

Figure. \ref{fig:test_probability} demonstrates the success probabilities of the experiments. 
The success probabilities for Grover's search (GSA), the partial Grover's search (No QMP), and our new QMP search algorithm (QMP) are described in Eq. (\ref{eq:succ_prob_GS}), (\ref{eq:succ_prob_partial}), and (\ref{eq:QMPGS_succ_prob_reform}), respectively.
This histogram shows that the QMP outperforms the GSA and the No QMP.

The theoretical success probabilities of D5M5 and D5D5M5 are $25.8 \%$ and $60.2 \%$, respectively (Eq. (\ref{eq:succ_prob_GS})). On the other hand, because of the quantum noise, their measured success probabilities ($2.3 \%$ and $3.1 \%$, respectively as shown in Fig. \ref{fig:test_probability}) of them are much lower than the theoretical probabilities.
The No QMP circuits of G2D3M3 and G3D2M2 have $19.5 \%$ and $12.5 \%$ theoretical success probabilities, respectively (Eq. (\ref{eq:succ_prob_partial})).
As shown in Fig. \ref{fig:ex_probability}, the No QMP circuits have a lower success probability when they have more blocks.
On the other hand, the QMP circuits of G2D3M3 and G3D2M2 have $78.1 \%$ and $100 \%$ theoretical success probabilities, respectively (Eq. (\ref{eq:QMPGS_succ_prob_reform})).
This is consistent with Fig. \ref{fig:ex_probability} that the success probabilities increase when they have more blocks.
As shown in Fig. \ref{fig:ex_probability}, QMP results outperform the GSA and the No QMP circuits regardless of qubit mapping manners (default or manual).
The main reason for this is that the QMP circuits have greater theoretical success probabilities than the others as shown in Fig. \ref{fig:ex_probability}

In the comparison between G2D3M3 and G3D2M2 on $\textit{IBM\_Washington}$, the QMP circuits of G3D2M2 have much greater success probabilities than G2D3M3 and the former has shorter circuit depths than the latter as shown in Fig. \ref{fig:test_eqcv}. Therefore, the QMP circuits show a big increase in success probabilities from G2D3M3 ($12.0 \%$ for QMP (default) and $20.1 \%$ for QMP (manual)) to G3D2M2 ($41.6 \%$ for QMP (default) and $70.1 \%$ for QMP (manual)).

In the comparison between $\textit{IBM\_Washington}$ and $\textit{IBMQ\_Brooklyn}$, the results from $\textit{IBM\_Washington}$ show a little bit better results than the results from $\textit{IBMQ\_Brooklyn}$ as shown in the middle of Fig. \ref{fig:test_probability}.
We think this is because $\textit{IBM\_Washington}$ has a greater quantum volume ($64$) than the quantum volume ($32$) of $\textit{IBMQ\_Brooklyn}$.

In the comparison between QMP implementations, QMP (manual) outperforms QMP (default) as shown in Fig. \ref{fig:test_probability}.
The manual mapping gets benefits not only from a buffer qubit to mitigate crosstalk between circuits but also from shorter circuit depth.
As shown in Fig. \ref{fig:G2D3M3_brooklyn_layout},  \ref{fig:G2D3M3_washington_layout}, and  \ref{fig:G3D2M2_washington_layout} in  Appendix \ref{app:no_qmp_orig_grover_layout}, QMP (default) circuits have $1.7 \sim 2.3$ times longer circuit depths.
QMP (default) does not guarantee for logical qubits in a circuit to be always allocated adjacent physically due to variance of the calibration data.
For example, the logical qubits (from $0$ to $5$) of the first circuit of the QMP circuit are mapped to separated physical qubits (nine qubit distance between qubit $3$ and qubit $5$) as shown in Fig. \ref{fig:G2D3M3_brooklyn_layout} (a).
This kind of physical mapping separation in one circuit is observed in all QMP (default) in Fig. \ref{fig:G2D3M3_brooklyn_layout} (a),  \ref{fig:G2D3M3_washington_layout} (a), and  \ref{fig:G3D2M2_washington_layout} (a).
Even worse is the qubit separation by the qubits of other circuits.
In Fig. \ref{fig:G3D2M2_washington_layout} (a), qubit $0$ is in the middle of qubits $12$ and $14$ which are qubits for the third circuit.
Those qubit separations require a lot of SWAP gates to connect them.
On the other hand, QMP (manual) is designed to avoid those problems as well as  better connectivity, and fewer errors of qubits and connections.
This shows how crucial the initial qubit mapping is on NISQ computers.
For more efficient and systematic qubit mapping, advanced qubit mapping algorithms \cite{das2019case, liu2021qucloud, niu2022parallel, niu2021enabling} can be applied to our QMP algorithm.

These experimental results are also compared with the results from trapped ion quantum computers such as Honeywell Quantum Computer (HQC) ($49.0 \%$) and IonQ quantum computer ($9.3 \%$) (dotted red lines).
The HQC and the IonQ quantum computer results are from $D5D5M5$ of the GSA and $G3D2M2$ of the No QMP, respectively in Ref. \cite{zhang2022quantum}.
These cases are chosen because they are the highest success probabilities out of various experimental settings with the same problem (one target out of $32$) in Ref. \cite{zhang2022quantum}.
The same circuit implementation is used for the HQC and the IonQ quantum computer cases as described in Sec. \ref{sec:implementation}.
Figure \ref{fig:test_probability} shows how much the QMP improves the success probability compared to trapped ion quantum computers.
The main reason the QMP of G3D2M2 has a higher success probability than the probability of HQC D5D5M5 is that the former has a greater theoretical success probability than the latter despite the higher quantum volume $1024$ of the latter.
This result shows a way to redeem the relatively low quantum volume of superconducting quantum computers by using more qubits.

\subsubsection{Expected quantum circuit volume}

\begin{figure}[t!]
\centering
\includegraphics[width=0.95\textwidth]{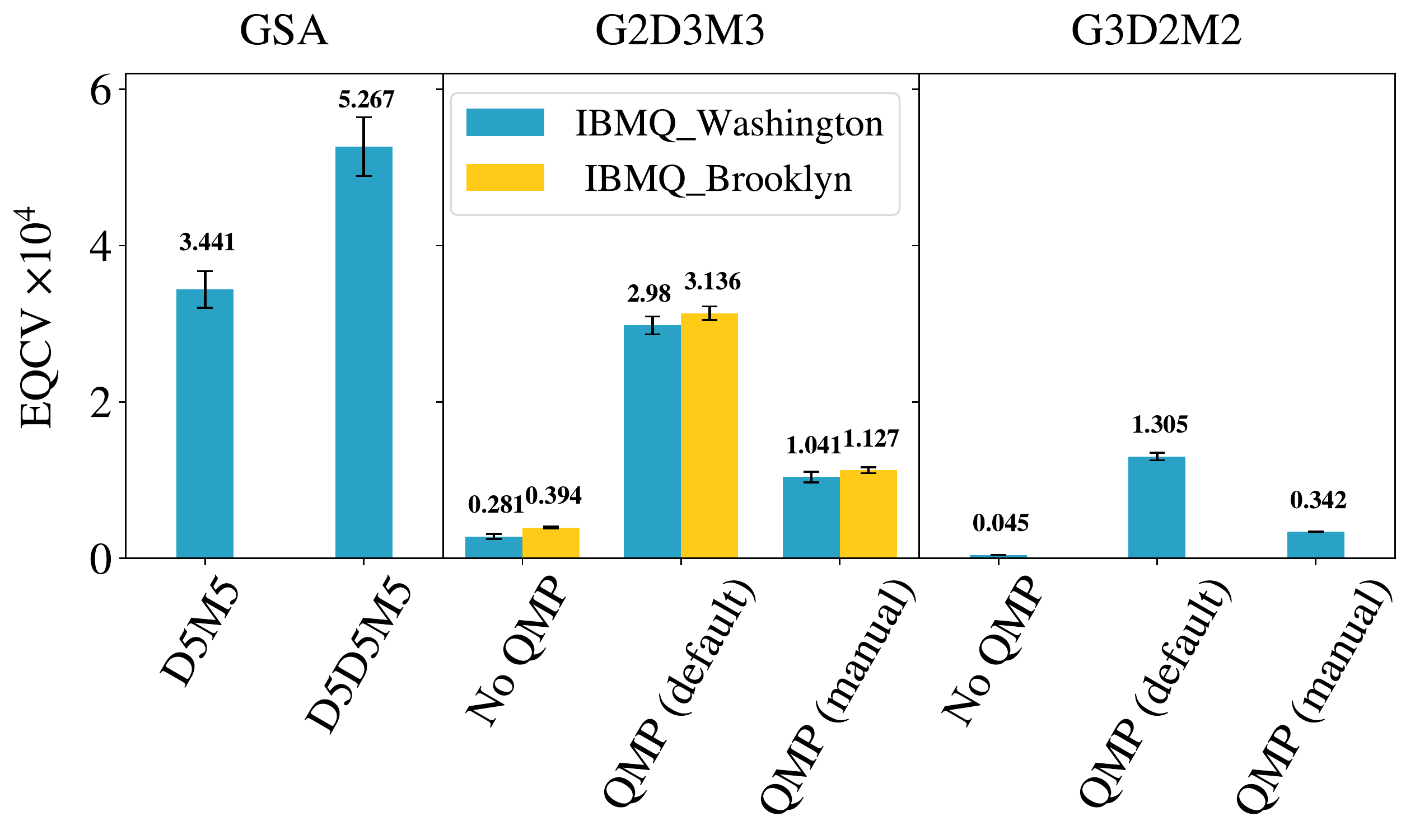}
\caption{EQCV of the experimental results. Refer to Sec. \ref{sec:implementation} and Sec. \ref{sec:setting} for each experimental case. The experiments were performed with $8192$ trials (shots) and 30 running repetitions, and we used the average value of the $30$ repetitions for the height of the histograms with the standard deviation error bar.}
\label{fig:test_eqcv}
\end{figure}

Figure \ref{fig:test_eqcv} shows each case's Expected Quantum Circuit Volume (EQCV).
EQCV is a criterion to measure circuit implementation efficiency with respect to the success probability.
The smaller EQCV is the more efficient circuit implementation on a quantum device.
The G3D2M2 circuits have the shortest circuit depths than the other circuits and the highest success probability even though they have two times and eight times more qubits than G2D3M3 and the GSA, respectively.
Therefore, the G3D2M2 circuits show the best EQCV.
On the other hand, the GSA circuits have the longest circuit depths and the lowest success probability even though they have four times and eight times a smaller number of qubits than G2D3M3 and G3D2M2, respectively. Hence, they show the worst EQCV.
The circuit depth and the qubit mapping layout of each experiment case are listed in Appendix  \ref{app:no_qmp_orig_grover_layout}.
In the GAS cases, even though D5D5M5 has a little bit greater success probability than D5MD, the circuit depth of the former is much longer than the depth of the latter.
So, the EQCV of D5M5 is smaller than the EQCV of D5D5M5.
In the G2D3M3 comparison, each case has similar circuit depths on both $\textit{IBM\_Washington}$ and $\textit{IBMQ\_Brooklyn}$.

\section{Conclusion and perspective}
\label{sec:conclusion}
We propose a QMP Grover search algorithm.
This new algorithm has a bigger rotation angle than the angle of the canonical Grover operator as shown in Eq. (\ref{eq:QMPGS_angle}).
Also, we showed that this new algorithm is more efficient than the canonical Grover's algorithm on IBM quantum computers and we expect that this advantage will be applied to other superconducting quantum computers.
In the case of multiple targets, since the success probability of each block depends on the number of targets, that is, the distribution of targets, it is possible to apply a different number of partial Grover operators per block.
This accompanies different circuit depths per block, and this is also allowed in QMP.
The different target densities in blocks (circuits) in the QMP will be studied in our future work.
In this study, even though we adapted a simple qubit mapping method (QMP (manual)), systematic qubit mapping algorithms are studied in Ref. \cite{das2019case, liu2021qucloud, niu2022parallel, niu2021enabling}. These qubit mapping algorithms will be useful to implement more complicated QMP circuits in a bigger quantum system. 
Also, recently developed error mitigation methods, such as crosstalk mitigation  \cite{ ohkura2022simultaneous,murali2020software}, hardware-aware compiling \cite{niu2021enabling}, and qubit partitioning \cite{niu2021enabling, niu2022parallel}, will be studied and implemented with our new QMP algorithm for our future work.

Our new quantum algorithm shows a way to redeem the relatively low quantum volume of superconducting quantum computers by using more qubits.
We utilize the fact that superconducting quantum computers have much more qubits than quantum volumes.
Also, the limited connectivity of superconducting quantum computers can be an advantage to applying QMP because it reduces crosstalk between each circuit in QMP.

Even though QMP has more advantages in superconducting quantum computers, it can be applied to trapped ion quantum computers. The number of circuits running concurrently is limited due to the relatively small number of qubits.
However, this QMP Grover's search will have a shorter circuit depth than the canonical Grover's search and this will increase the accuracy of the result.

From an algorithmic point of view, applying QMP to a quantum algorithm is not straightforward because it needs a decomposition of the algorithm.
In Grover's search case, we applied partial Grover's search \cite{groverTradeoffsQuantumSearch2002}.
However, if a QMP of a quantum algorithm reduces the circuit depth, then the QMP algorithm will have many benefits from the shorter circuit depth on NISQ devices.
Therefore, it will be valuable to develop QMP algorithms for quantum algorithms.

The number of blocks in the QMP is limited by the number of qubits of the target quantum processor and the search problem size although the efficiency of the QMP circuit increases as a more number of blocks are used as shown in Fig. \ref{fig:ex_probability}.
Therefore, the usage of this algorithm is limited when quantum processors have a small number of qubits like $\it IBMQ\_Kolkata$, $\it IBMQ\_Montreal$, $\it IBMQ\_Mumbai$.
However, this QMP algorithm will be more useful as the number of qubits of quantum processes increases to more than one thousand.

\section*{Acknowledgments}

We would like to thank the Brookhaven National Laboratory operated IBM-Q Hub.
This research used quantum computing resources of the Oak Ridge Leadership Computing Facility, which is a DOE Office of Science User Facility supported under Contract DE-AC05-00OR22725. 
This research used resources of the National Energy Research Scientific Computing Center, a DOE Office of Science User Facility supported by the Office of Science of the U.S. Department of Energy under Contract No. DE-AC02-05CH11231 using NERSC award DDR-ERCAP0022229.
We acknowledge the access to IonQ and Honeywell Quantum Solution through the Microsoft Azure Quantum grant No. 35984. 
The work of Vladimir Korepin was supported by the U.S. Department of Energy, Office of Science, National Quantum Information Science Research Centers, Co-design Center for Quantum Advantage (C2QA) under Contract No. DE-SC0012704.

\section*{Data Availability}

All codes and the datasets generated during and/or analyzed during the current study are available in the GitHub repository, https://github.com/yukwangmin/QMP\_GS.


\begin{appendices}

\section{Partial Measurement}
\label{app:meas}

\begin{lstlisting}[language={Python}, caption={A sample code for partial measurement extraction from the whole measurement}, label={list:after}]
import numpy as np

# count : return of result.get_counts()
# N     : The total number of bits of the whole measurement
# least : The least significant bit index of the partial measurement
# n     : The number of bits for partial measurement
def partitionMeasurement(count, N, least, n):
    
    cnt = {}

    
    for i in range(2**n):
        key = f'{{0:0>{n}b}}'.format(i)
        cnt[key] = 0
    
    for j in range(2**n):
        key = f'{{0:0>{n}b}}'.format(j)
    
        for i in range(2**(N-n)):
            bin = f'{{0:0>{N-n}b}}'.format(i)
            nkey = bin[:N-n-least] + key + bin[N-n-least:]
            cnt[key] += count.get(nkey, 0)
            
    return cnt   


\end{lstlisting}

\section{Qubit mapping layout and circuit depth}
\label{app:no_qmp_orig_grover_layout}

\renewcommand{\thefigure}{B\arabic{figure}}
\renewcommand{\theHfigure}{B\arabic{figure}}
\setcounter{figure}{0}

In this section, the layout of IBM quantum computers and qubit mappings are visualized. Each sub-figures are results of qiskit.visualization.plot\_circuit\_layout() function calls in Qiskit and the numbers in the black circles represent the logical qubits of the QMP circuit before the transpiling.

\begin{figure}[h]
\centering
  \subfloat[QMP (default) layout (depth: 154)]{%
    \includegraphics[width=0.4\textwidth]{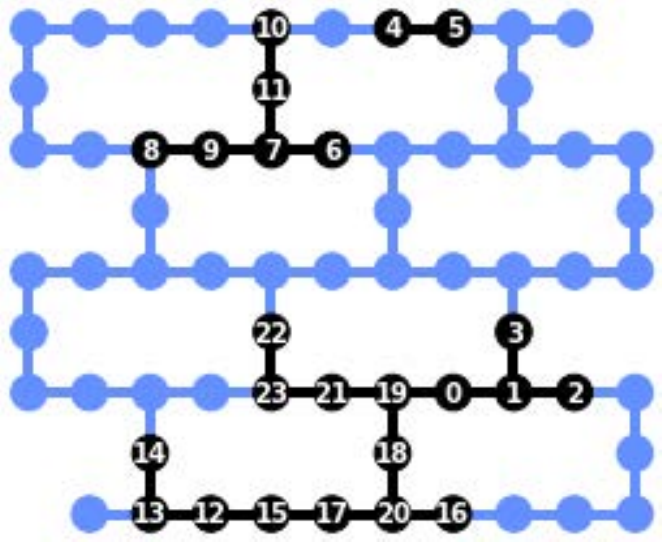}}
  \qquad    
  \subfloat[QMP (manual) layout (depth: 86)]{%
    \includegraphics[width=0.4\textwidth]{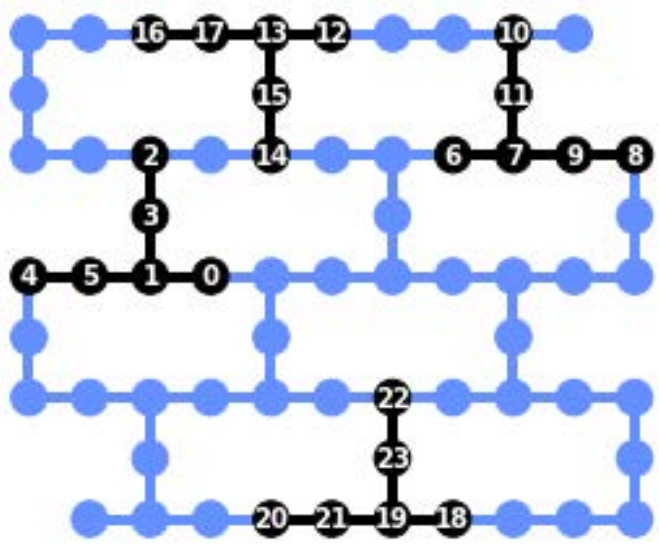}}~
  \vspace{0.5cm}
  \caption{Circuit mapping layouts for G2D3M3 on IBMQ\_Brooklyn.}
  \label{fig:G2D3M3_brooklyn_layout}
\end{figure}

\begin{figure}[h]
\centering
  \subfloat[QMP (default) layout (depth: 149)]{%
    \includegraphics[width=0.45\textwidth, height=0.35\textwidth]{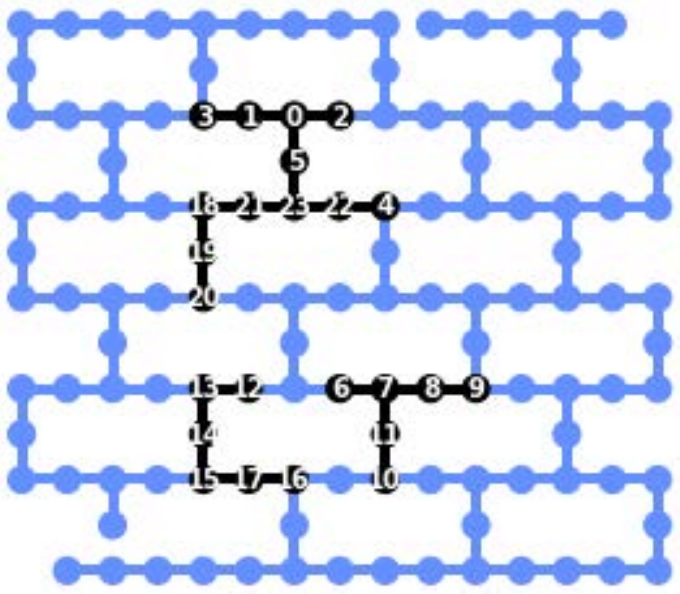}}
  \qquad
  \subfloat[QMP (manual) layout (depth: 87)]{%
    \includegraphics[width=0.45\textwidth, height=0.35\textwidth]{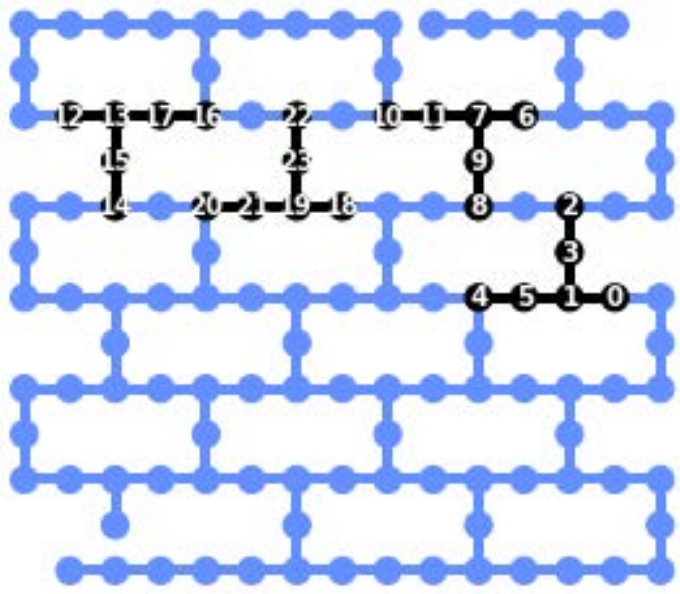}}
  \vspace{0.5cm}
  \caption{Circuit mapping layouts for G2D3M3 on IBM\_Washington.}
  \label{fig:G2D3M3_washington_layout}
\end{figure}

\begin{figure}[h]
\centering
  \subfloat[QMP (default) layout (depth: 113)]{%
    \includegraphics[width=0.45\textwidth]{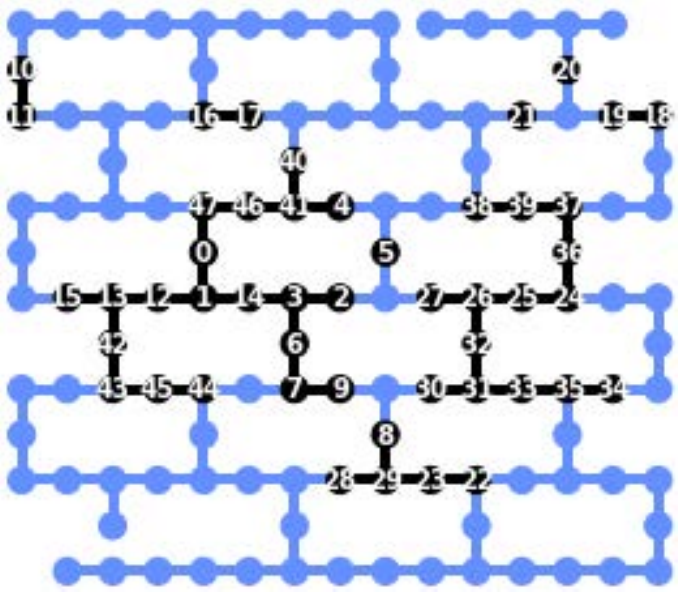}}
  \qquad
  \subfloat[QMP (manual) layout (depth: 50)]{%
    \includegraphics[width=0.45\textwidth]{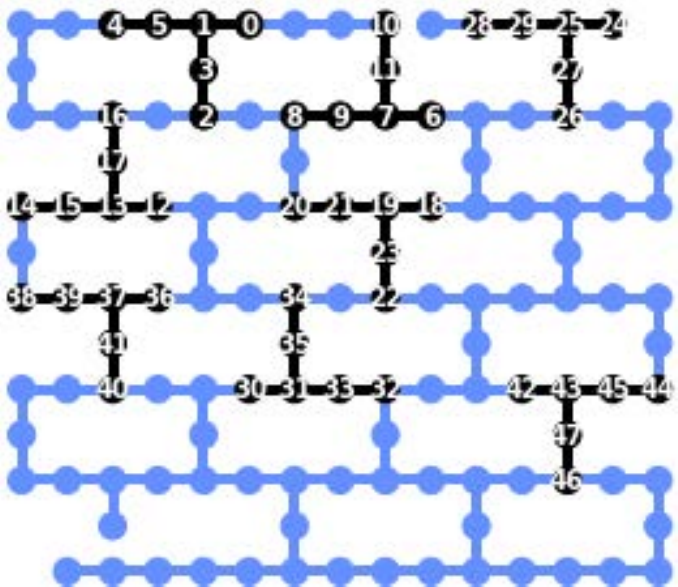}}
  \vspace{0.5cm}
  \caption{Circuit mapping layouts for G3D2M2 on IBM\_Washington.}
  \label{fig:G3D2M2_washington_layout}
\end{figure}

\end{appendices}


\clearpage



\bibliography{refs}

\end{document}